\documentclass[10pt,aps,prb,twocolumn,amsmath,amssymb,superscriptaddress,nobibnotes]{revtex4-1}
\usepackage{amsmath}
\usepackage{graphicx}
\usepackage{color}
\usepackage{multirow}
\usepackage[caption = false]{subfig}
\usepackage{array}
\usepackage{threeparttable}
\usepackage{mhchem}
\usepackage{mathtools}
\usepackage{siunitx}
\usepackage{centernot}
\usepackage{tikz}
\tikzset{font={\fontsize{11pt}{12}\selectfont}}
\usetikzlibrary{shapes,arrows}
\usepackage{lipsum}
\usepackage[normalem]{ulem}
\makeatletter

\renewcommand*{\p@subsection}{}

\renewcommand*{\p@subsubsection}{}
\makeatother

\usepackage{hyperref}
\hypersetup{
	linkcolor=blue,          
	citecolor=blue,        
	urlcolor=blue,           
}

\newcommand{\ket}[1]{|#1\rangle}

\newcommand{\inner}[2]{\langle#1|#2\rangle}

\newcommand{\expect}[3]{\langle#1|\hat{#2}|#3\rangle}
\newcommand{\expecth}[3]{\langle#1|#2|#3\rangle}

\begin{document}
\title{Response properties in phaseless auxiliary field quantum Monte Carlo}

\author{Ankit Mahajan}
\email{ankitmahajan76@gmail.com}
\affiliation{Department of Chemistry, Columbia University, New York, NY 10027, USA}
\affiliation{Department of Chemistry, University of Colorado, Boulder, CO 80302, USA}

\author{Jo S. Kurian}
\affiliation{Department of Chemistry, University of Colorado, Boulder, CO 80302, USA}

\author{Joonho Lee}
\affiliation{Department of Chemistry, Columbia University, New York, NY 10027, USA}
\affiliation{Department of Chemistry and Chemical Biology, Harvard University, Cambridge, MA 02138, USA}

\author{David R. Reichman}
\affiliation{Department of Chemistry, Columbia University, New York, NY 10027, USA}

\author{Sandeep Sharma}
\email{sanshar@gmail.com}
\affiliation{Department of Chemistry, University of Colorado, Boulder, CO 80302, USA}

\begin{abstract}
We present a method for calculating first-order response properties in phaseless auxiliary field quantum Monte Carlo (AFQMC) through the application of automatic differentiation (AD). Biases and statistical efficiency of the resulting estimators are discussed. Our approach demonstrates that AD enables the calculation of reduced density matrices (RDMs) with the same computational cost scaling as energy calculations, accompanied by a cost prefactor of less than four in our numerical calculations. We investigate the role of self-consistency and trial orbital choice in property calculations. 
\end{abstract}
\maketitle

\section{Introduction}
The calculation of energy gradients is of great importance in electronic structure theory. Derivatives of the energy with respect to external perturbations can be used to evaluate many properties of interest. Nuclear gradients are used for geometry optimization and molecular dynamics simulations. Parameter derivatives are useful in wave function and basis set optimizations. Throughout the history of quantum chemistry, the development of efficient methods for evaluating energy gradients has thus been a focal point due to their central importance.\cite{pulay1969ab,pople1979derivative,handy1984evaluation,salter1989analytic,helgaker2012recent} This development often involved meticulous derivation of analytic expressions and manual coding of implementations. The development of efficient algorithms has led to cost-effective gradient evaluations, comparable in cost to the evaluation of the energy. In recent years, automatic or algorithmic differentiation (AD) has gained recognition as a powerful tool for calculating derivatives of functions defined by computer programs, notably in machine learning. By applying the chain rule to elementary steps in a computation, AD can calculate numerically exact derivatives, considerably reducing the implementation effort.\cite{griewank2008evaluating} The reverse-mode AD is particularly remarkable, enabling the evaluation of first-order derivatives with respect to an arbitrary number of parameters at the same cost scaling as the underlying function. As a result, it becomes an appealing choice for calculating all nuclear forces or reduced density matrices in quantum chemistry. Various studies have explored the utility of AD in conventional electronic structure methods.\cite{ekström2010arbitrary,bast2011ab,abbott2021arbitrary,zhang2022differentiable,kasim2022dqc} Reference \citenum{liao2019differentiable} showcased the effectiveness of AD in differentiating tensor network-based functions.

Quantum Monte Carlo (QMC) approaches comprise a scalable and accurate suite of alternatives to traditional quantum chemistry methods.\cite{kalos1974helium,ceperley1977monte,ceperley1980ground,nightingale1998quantum,foulkes2001quantum,becca2017quantum} Among these, the auxiliary field quantum Monte Carlo (AFQMC) approach stands out as a powerful tool for investigating correlated electronic systems.\cite{zhang1995constrained,zhang1997constrained,zhang2003quantum} It has been used to study a wide variety of systems, including solids and molecules with a range of correlation strengths.\cite{al2006auxiliary,suewattana2007phaseless,hao2018accurate,lee2020utilizing,malone2022ipie,shee2019singlet} However, the evaluation of energy derivatives has posed challenges for projection QMC methods. Mixed estimators are usually used in energy calculations with these methods, which are not accurate for general properties or nuclear gradients. An incorrect application of the Hellman-Feynman theorem leads to this mixed estimator that is easy to evaluate but leads to an uncontrolled bias for operators that do not commute with the Hamiltonian. Although various approximations can address this shortcoming,\cite{badinski2010methods,casalegno2003computing,filippi2000correlated,moroni2014practical,assaraf2003zero,motta2017computation,thomas2015analytic} many challenges still remain due to uncontrolled biases or large variances. It is worth mentioning that within the context of variational Monte Carlo (VMC), AD was first employed by Sorella and Capriotti to calculate nuclear gradients.\cite{sorella2010algorithmic} Additionally, there have been studies of AD in diffusion Monte Carlo (DMC)\cite{poole2014calculating} and lattice VMC\cite{zhang2019automatic}, following an approach similar to that presented here.

The problem of calculating derivatives of complicated stochastic functions plays a crucial role in various domains ranging from sensitivity analysis in finance\cite{glasserman2004monte} and queueing theory in management\cite{glynn1990likelihood} to optimizations in reinforcement learning.\cite{bengio2013estimating} An effective approach for formulating low-variance estimators involves utilizing correlated sampling, by employing common random numbers for example, to achieve small cost differences with high accuracy. When Monte Carlo simulations involve discrete random variables, like branching in QMC calculations, biases can arise in derivative estimators, and addressing these errors is an active area of research.\cite{mohamed2020monte} In this paper, we leverage some of this technology to compute first-order response properties in AFQMC. Specifically, we demonstrate that reverse-mode AD enables the calculation of Reduced Density Matrices (RDMs) with the same cost scaling as energy calculations. 

This paper is organized as follows: we first review relevant parts of the AFQMC algorithm (Section \ref{sec:afqmc}) and existing approaches for calculating properties (Section \ref{sec:properties}). Then we discuss use of AD in AFQMC, including various contributions to the derivatives as well as biases in our method (Section \ref{sec:derivatives}). We then present illustrative numerical results to demonstrate the accuracy and statistical efficiency of our method (Section \ref{sec:illustrative}). Finally, we compare AD-AFQMC dipole moments of some molecules with other quantum chemistry methods and experiments (Section \ref{sec:results}).

\section{Theory}\label{sec:theory}
\subsection{Review of AFQMC}\label{sec:afqmc}
In this section, we present an outline of the AFQMC algorithm, focusing on the aspects relevant to gradient calculations. We refer the reader to recent review articles \cite{motta2018ab,shi2021some,lee2022twenty} for a more comprehensive description. In AFQMC, the ground state is represented as a weighted sum of non-orthogonal Slater determinants
\begin{equation}
   \ket{\Psi_0} = \sum_i w_i \frac{\ket{\phi_i}}{\inner{\psi_T}{\phi_i}},
\end{equation}
where \(w_i\) are weights, \(\ket{\phi_i}\) are walker Slater determinants with complex orbitals, and \(\ket{\psi_T}\) is the trial state. This state is obtained by applying an exponential form of projector onto the trial state as
\begin{equation}
   e^{-\tau \hat{H}}\ket{\psi_T} \xrightarrow{\tau\rightarrow\infty}\ket{\Psi_0},
\end{equation}
where \(\tau\) is the imaginary time, and we assume \(\inner{\psi_T}{\psi_0}\neq 0\). We use the quantum chemistry Hamiltonian given by
\begin{equation}
	\hat{H} = \sum_{ij} h_{ij}\hat{a}_i^{\dagger}\hat{a}_j + \frac{1}{2}\sum_{\gamma}\left(\sum_{ij}L^{\gamma}_{ij}\hat{a}_i^{\dagger}\hat{a}_j\right)^2,\label{eq:ham}
\end{equation}
where \(h_{ij}\) are one-electron integrals and \(L^{\gamma}_{ij}\) are Cholesky decomposed two-electron integrals in an orthonormal orbital basis. To sample the action of the propagator, the Hubbard-Stratonovic transform is used to write the short-time propagator as
\begin{equation}
   e^{-\Delta\tau\hat{H}} = \int d \mathbf{x} p(\mathbf{x})\hat{B}(\mathbf{x}),\label{eq:prop}
\end{equation}
where \(\mathbf{x}\) is the vector of auxiliary fields, \(p(\mathbf{x})\) is the standard normal distribution of the auxiliary fields, and \(\hat{B}(\mathbf{x})\) is a complex propagator given by the exponential of a one-body operator. Due to Thouless' theorem, \(\hat{B}(\mathbf{x})\) acts on a Slater determinant \(\ket{\phi}\) as
\begin{equation}
   \hat{B}(\mathbf{x})\ket{\phi} = \ket{\phi(\mathbf{x})},
\end{equation}
where \(\ket{\phi(\mathbf{x})}\) is another Slater determinant obtained by a linear transformation of the orbitals in \(\ket{\phi}\). In the hybrid approximation, importance sampling is achieved by shifting the auxiliary fields with a force bias given as
\begin{equation}
   \bar{x}_{\gamma} = -\sqrt{\Delta\tau} \frac{\expecth{\psi_T}{\sum_{ij}L^{\gamma}_{ij}\hat{a}_i^{\dagger}\hat{a}_j}{\phi}}{\inner{\psi_T}{\phi}},\label{eq:fb}
\end{equation}
and with the weights propagated as
\begin{equation}
   w_i(\mathbf{x}) = \bigg\vert \frac{\inner{\psi_T}{\phi(\mathbf{x})}}{\inner{\psi_T}{\phi}} e^{\mathbf{x}.\bar{\mathbf{x}} - \frac{\bar{\mathbf{x}}^2}{2}}\bigg\vert w_i.
\end{equation}
Force bias is not enough, by itself, to control the large fluctuations stemming from the sign or phase problem. The phaseless constraint\cite{zhang2003quantum} can be used to overcome the phase problem at the expense of a systematic bias in the sampled wave function by performing a cosine projection as
\begin{equation}
   w_i(\mathbf{x}) = \text{max}(0,\cos(\Delta\theta)),\label{eq:phaseless}w_i(\mathbf{x})
\end{equation} 
where
\begin{equation}
   \Delta\theta = \text{Arg}\left(\frac{\inner{\psi_T}{\phi(\mathbf{x})}}{\inner{\psi_T}{\phi}}\right).
\end{equation}
The size of this phaseless bias is dictated by the accuracy of the trial state. 

\begin{figure}
   \includegraphics[width=0.8\columnwidth]{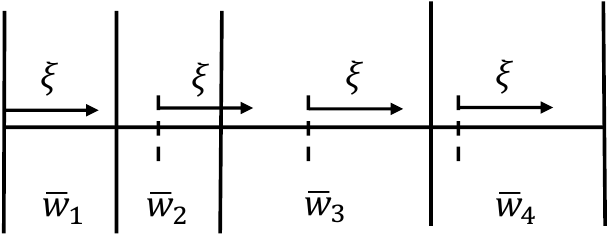}
   \caption{A schematic showing the stochastic reconfiguration (SR) procedure with four walkers. \(\xi\) is drawn from a uniform distribution on [0,\(\frac{1}{4}\)] and \(\bar{w}_i = \frac{w_i}{\sum_i w_i}\) denote normalized weights. In this instance, SR leads to the reconfiguration \(\left\{\phi_1,\phi_2,\phi_3,\phi_4\right\} \rightarrow \left\{\phi_1,\phi_3,\phi_3,\phi_4\right\}\).}\label{fig:sr}
\end{figure}

For statistical efficiency, we periodically perform reconfigurations of the walker population. We use the stochastic reconfiguration (SR) method\cite{buonaura1998numerical} where walkers with small weights are removed, and those with large weights are duplicated. Figure \ref{fig:sr} illustrates this procedure schematically. It has the desirable property that when the walker weights are uniform, the walker population remains unchanged. It is important to note that SR is a discontinuous function, as small changes in weights lead to different reconfiguration instances. These discontinuities are a source of bias in our AD gradient calculations, as described in section \ref{sec:derivatives}.

\subsection{Calculating properties in AFQMC}\label{sec:properties}
Given an observable \(\hat{O}\), the simplest way to estimate its expectation value in AFQMC is to use the mixed estimator given by
\begin{equation}
     \langle\hat{O}\rangle_{\text{mixed}} = \frac{\sum_{\phi} w_{\phi} \frac{\expecth{\psi_T}{\hat{O}}{\phi}}{\inner{\psi_T}{\phi}}}{\sum_{\phi} w_{\phi}}. \label{eq:mixed}     
\end{equation}
If \(\hat{O}\) does not commute with the Hamiltonian, this estimator has a bias due to its non-variational nature. One way to mitigate this error is to use the extrapolated estimator\cite{whitlock1979properties,ceperley1986quantum,rothstein2013survey}
\begin{equation}
   \langle\hat{O}\rangle_{\text{extrapolated}} = 2\langle\hat{O}\rangle_{\text{mixed}} - \frac{\expect{\psi_T}{O}{\psi_T}}{\inner{\psi_T}{\psi_T}}.
\end{equation}
The simplicity of evaluating this estimator comes with a crucial caveat; its accuracy heavily relies on the trial state employed. For instance, achieving convergence of dipole moments in reference \citenum{mahajan2022selected} required the use of a large multi-Slater trial state.

A more common approach to calculating properties in AFQMC is using backpropagation,\cite{zhang1995constrained} which overcomes the non-variationality issue by propagating the trial state backward as
\begin{equation}
   \langle\hat{O}\rangle_{\text{bp}} = \frac{\expecth{\psi_T}{e^{-\tau_{\text{bp}}\hat{H}}\hat{O}e^{-\tau\hat{H}}}{\psi_T}}{\expecth{\psi_T}{e^{-(\tau_{\text{bp}}+\tau)\hat{H}}}{\psi_T}}.
\end{equation}
A direct simulation of two independent random walks results in a highly noisy and computationally intensive estimator. To control this noise, importance sampling is employed to correlate the two walks. However, in certain systems, this procedure has been observed to introduce a significant phaseless bias. This bias can be reduced by undoing the phaseless approximation in the backpropagation, albeit at the expense of noisier estimates.\cite{motta2017computation} Backpropagation has also been used to evaluate nuclear gradients based on the Hellman-Feynman theorem.\cite{motta2018communication}

With the goal of bringing accuracy and computational cost of property calculations on par with that of calculating the energy, we pursue a response approach in this paper. Consider a perturbed Hamiltonian given by
\begin{equation}
   \hat{H}(\lambda) = \hat{H} + \lambda\hat{O},
\end{equation}
where \(\lambda\) is a real parameter. The response estimator can be written as
\begin{equation}
   \langle\hat{O}\rangle_{\text{response}} = \frac{dE(\lambda)}{d\lambda}\bigg|_{\lambda=0},
\end{equation}
where \(E(\lambda)\) is the AFQMC energy calculated with the perturbed Hamiltonian \(\hat{H}(\lambda)\). For a variationally optimized wave function, this estimator is equivalent to the expectation value by virtue of the Hellman-Feynman theorem. For non-variational methods like coupled cluster theory, response properties are found to be very accurate and are routinely used. One way to evaluate energy derivatives is using the method of finite differences. However, as AFQMC is a stochastic method, it is necessary to correlate the finite difference energy calculations to minimize the noise in the energy difference. Such correlated sampling methods have been used in AFQMC\cite{shee2017chemical} as well as other QMC methods\cite{assaraf2011chaotic,filippi2000correlated}. Using a common stream of random numbers is an intuitively appealing strategy to maintain two Monte Carlo runs close to each other, but technical challenges arise in ensuring that the runs remain coherent. A promising approach to address this challenge was recently reported in reference \citenum{chen2023algorithm}. A drawback of the correlated sampling finite difference method is the requirement for separate energy calculations for each perturbation, making the calculation of reduced density matrices and all nuclear forces computationally expensive, with the cost scaling linearly with the number of perturbations.  In the next section, we will explore how reverse-mode automatic differentiation utilizes the common random number strategy, while also allowing calculations of responses to multiple perturbations at a cost comparable to that of a single energy calculation.

\subsection{Derivatives of AFQMC energy}\label{sec:derivatives}
We can write the AFQMC energy expression making the dependence on the perturbation explicit as
\begin{equation}
   \begin{split}
   E(\hat{H}_\lambda, \psi_T(\lambda), \left\{w_i(\lambda)\right\}, &\left\{\phi_i(\lambda)\right\})\\ 
   &= \frac{\sum_{i} w_{i}(\lambda) \frac{\expecth{\psi_T(\lambda)}{\hat{H}(\lambda)}{\phi_i(\lambda)}}{\inner{\psi_T(\lambda)}{\phi_i(\lambda)}}}{\sum_{i} w_{i}(\lambda)}.
   \end{split}
\end{equation}
Differentiating this expression with respect to \(\lambda\) gives
\begin{equation}
   \begin{split}
   \frac{dE}{d\lambda} =& \frac{\sum_{i} w_{i}(\lambda)\frac{\expecth{\psi_T(\lambda)}{\hat{O}}{\phi_i(\lambda)}}{\inner{\psi_T(\lambda)}{\phi_i(\lambda)}}}{\sum_{i} w_{i}(\lambda)} + \frac{\partial E}{\partial \psi_T(\lambda)}\frac{d\psi_T(\lambda)}{d\lambda}\\ 
    &+ \sum_{i} \left(\frac{\partial E}{\partial w_{i}(\lambda)}\frac{dw_i(\lambda)}{d\lambda} + \frac{\partial E}{\partial \phi_{i}(\lambda)}\frac{d\phi_i(\lambda)}{d\lambda}\right).
   \end{split}\label{eq:dedl}
\end{equation}
The first term is the mixed estimator \ref{eq:mixed}. The second term contains the explicit effect of the relaxation of the trial state in response to the perturbation on the energy. It also appears in the response calculation of many conventional quantum chemistry methods where it is evaluated (implicitly) using the coupled perturbed Hartree Fock (CPHF) equations.\cite{gerratt1968force} The final term contains contributions from the derivatives of weights and walkers, which, in turn, are influenced by the response of the trial state. The dependence of weights and walkers on the perturbation is accumulated through many steps of AFQMC propagation and local energy measurements. One can evaluate their derivatives by repeated application of the chain rule. AD offers an efficient and convenient way to accomplish this task.

All these contributions can be evaluated using AD on an energy function that incorporates the effects of the perturbation on the final energy. If stochastic reconfiguration is not performed, the AFQMC energy can be written as the integral
\begin{equation}
   E(\lambda) = \int d\mathbf{x} \mathcal{N}(\mathbf{x}) f(\mathbf{x}, \lambda),
\end{equation} 
where \(\mathbf{x}\) is the collection of normally distributed auxiliary fields for all time steps and walkers, and \(f\) denotes the AFQMC procedure, including propagation and measurement. For the sake of notational convenience, we have suppressed the dependence of \(f\) on other quantities that are not pertinent to the current discussion. The derivative of this integral can be obtained as
\begin{equation}
   \frac{dE}{d\lambda} = \int d\mathbf{x}\ \mathcal{N}(\mathbf{x}) \frac{\partial f}{\partial \lambda}(\mathbf{x}. \lambda),\label{eq:pathwise}
\end{equation}
Note that the phaseless constraint (\ref{eq:phaseless}) is not differentiable at the origin, but it is continuous and differentiable almost everywhere. The switching of derivative and integral is allowed in this case because all the operations performed in the evaluation of \(f\) are continuous and differentiable almost everywhere (within the region of interest).\cite{glasserman2004monte} As a result, ~(\ref{eq:pathwise}) provides a means to sample the derivative through Monte Carlo sampling, akin to the energy, and is known as a pathwise estimator. This estimator exhibits desirable properties of low variance, thanks to correlated sampling, and avoids numerical issues stemming from finite difference step sizes.  This pathwise estimator is precisely the one that is evaluated by performing automatic differentiation on the energy function. Upon introducing SR, the energy can be written as
\begin{equation}
   E(\lambda) = \int d\mathbf{x}d\boldsymbol{\xi}\  \mathcal{N}(\mathbf{x}) \mathcal{U}(\boldsymbol{\xi}) \tilde{f}(\mathbf{x}, \mathbf{\xi},\lambda),
\end{equation}
where \(\mathbf{\xi}\) are uniformly distributed variables. Due to the discrete nature of SR, the function \(\tilde{f}\) is not continuous and the pathwise estimator is biased. Several methods have been developed to address biases arising from discontinuities, score function or likelihood ratio estimators being a popular choice.\cite{glynn1990likelihood,glasserman2004monte,schulman2015gradient} However, in preliminary numerical experiments, we observed this method to be quite noisy. Hence, we defer a detailed study of these bias mitigation techniques to future work. In this paper, we use the biased pathwise estimator to sample the derivative of the energy function. We present a numerical analysis of this bias in section \ref{sec:results}.

\subsection{Implementation details}
In AFQMC energy calculations, a sequence of phaseless imaginary time propagation steps is executed before reconfiguration and energy measurement. At the end of this block of steps, the walker weights and local energies are averaged to yield estimates of the energy along with its associated stochastic error. To obtain energy derivative samples, AD is performed on the averaged energy from a sufficiently large number of such blocks. The number of blocks to be included in the AD calculation is dictated by the time required for the convergence of walker and weight derivatives (\ref{eq:dedl}).  We present a numerical investigation of this convergence in section \ref{sec:illustrative}.

AD can be used in two modes: forward and reverse. In forward-mode AD, the derivative of the energy function is evaluated by propagating the perturbation through the energy function. This approach calculates the response to one perturbation at a time, resulting in a computational cost scaling comparable to correlated finite difference methods. However, it offers the advantage of having no finite difference errors and requires low memory, making it suitable for evaluating a few properties in large systems. For example, one can evaluate the dipole moment of a system by using the perturbation given by
\begin{equation}
   \hat{H}(\boldsymbol{\varepsilon}) = \hat{H} + \boldsymbol{\varepsilon}.\hat{\mathbf{d}},
\end{equation}
where \(\hat{\mathbf{d}}\) is the dipole operator, and calculating the derivative with respect to the electric field \(\boldsymbol{\varepsilon}\), requiring at most three forward-mode AD calculations. In reverse-mode AD, gradients are computed by propagating changes in energy backward through to the perturbations. Remarkably, it shares the same computational cost scaling as the energy function itself when calculating gradients with respect to an arbitrary number of perturbations. Depending on the specific implementation, the cost is usually between 3 to 5 times that of the energy evaluation. This property makes reverse-mode AD well-suited for efficiently calculating many properties simultaneously. For example, one can evaluate one and two-particle reduced density matrices (RDM) by using the following perturbed Hamiltonian
\begin{equation}
   \hat{H}(\lambda) = \hat{H} + \sum_{ij} \lambda_{ij}\hat{a}_i^{\dagger}\hat{a}_j + \sum_{iajb} \gamma_{ipjq}\hat{a}_i^{\dagger}\hat{a}_j^{\dagger}\hat{a}_q\hat{a}_p,
\end{equation}
and evaluating derivatives with respect to \(\lambda_{ij}\) and \(\gamma_{ipjq}\), respectively. 

Reverse-mode AD achieves its impressive computational cost scaling by storing intermediates during the evaluation of the energy function resulting in a linear increase in memory usage with the number of AFQMC time steps to be differentiated. Since the derivatives of weights and walkers need to be converged with respect to the propagation time, this leads to a substantial memory cost for reverse-mode AD with increasing system size. We use checkpointing,\cite{griewank2008evaluating} which recalculates some intermediates on the backward pass instead of storing them, to reduce this cost. This reduction in memory usage comes at the expense of an increased computational cost prefactor.  

For capturing the orbital relaxation of a HF trial state, the AFQMC energy function can be written to include a call to a HF solver. We use a fixed and conservatively large number of SCF iterations to ensure that the derivatives are fully converged. Additionally, issues related to degeneracies in reverse-mode AD of eigendecompositions can be dealt with by ignoring rotations in degenerate spaces.\cite{zhang2022differentiable,liao2019differentiable}

\section{Results}\label{sec:results}
In this section, we present the results of our response ph-AFQMC calculations, referred to as AD-AFQMC, and analyze both systematic and stochastic errors. Our analysis focuses on the accuracy of 1-RDM and dipole moment calculations using AFQMC with a HF trial state. We used PySCF\cite{sun2018pyscf} to obtain molecular integrals and to perform quantum chemistry wave function calculations. The near-exact Density Matrix Renormalization Group (DMRG) calculations on hydrogen chains were done using the block2 code.\cite{zhai2021low} We used QMCPACK\cite{kim2018qmcpack} to do backpropagation AFQMC calculation. The code used to perform AD-AFQMC calculations is available in a public repository\cite{dqmc_code}, with the Jax\cite{jax2018github} library used for AD. Reverse-mode AD was employed for 1-RDM calculations, while forward-mode AD was used for dipole moments (except for self-consistent calculations). Input and output files for all calculations can be accessed from a public repository.\cite{afqmc_files} We used a time step of 0.01 a.u. in all ph-AFQMC calculations. Cholesky decompositions were calculated up to a threshold error of \(10^{-5}\). Errors due to these approximations can be controlled systematically and we estimate them to be smaller than the statistical errors in the results presented here. 

\subsection{Illustrative results}\label{sec:illustrative}
\begin{figure}
   \includegraphics[width=\columnwidth]{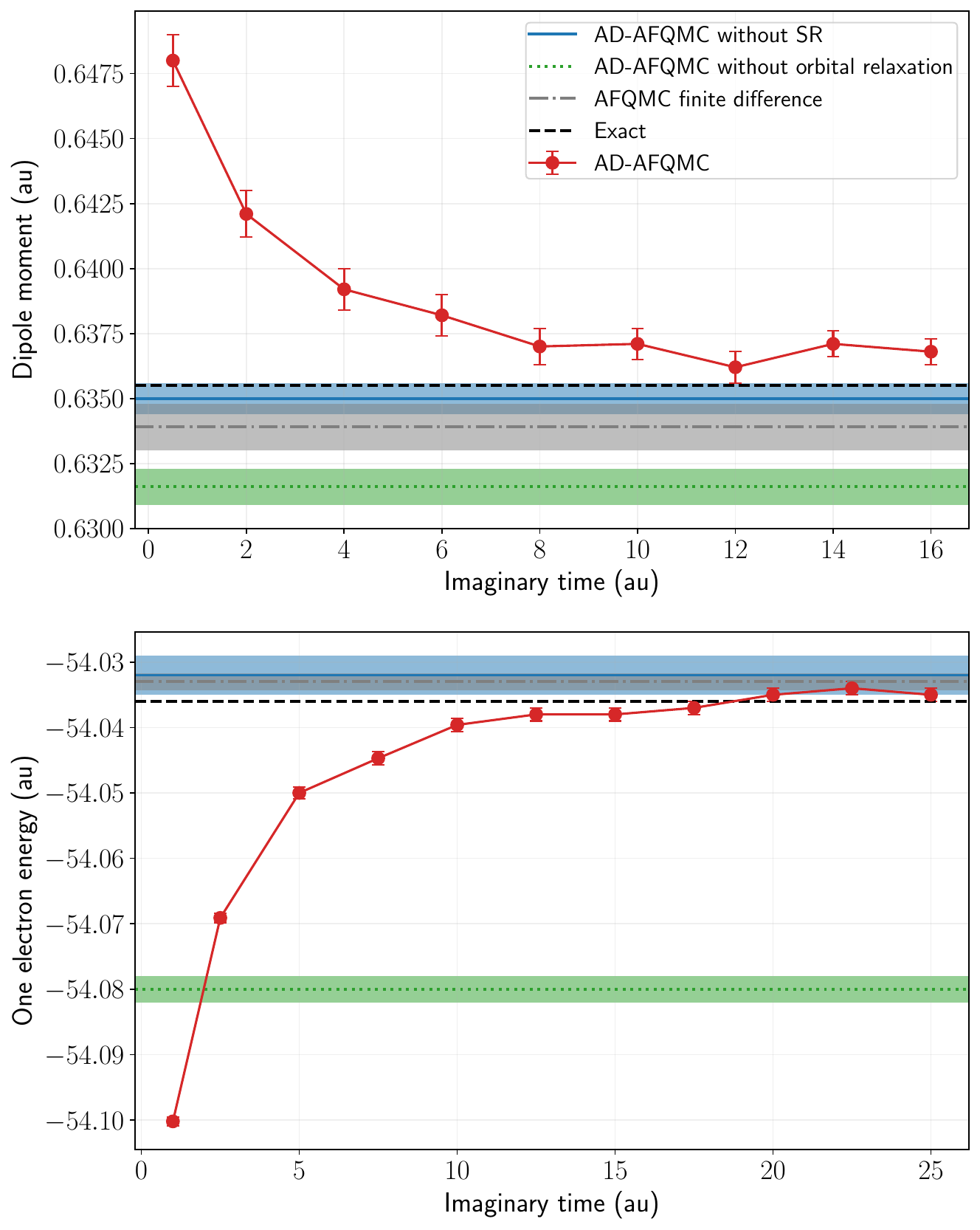}
   \caption{Dipole moment of \ce{NH3} in the cc-pVDZ basis (top) and one electron energy of a \ce{H20} chain in the STO-6G basis (bottom) as a function of propagation time.}\label{fig:conv}
\end{figure}

First, we analyze various contributions to observables and their convergence with imaginary time. In the top panel of figure \ref{fig:conv}, we show the convergence of the AD-AFQMC dipole moment of the ammonia molecule in the cc-pVDZ basis at equilibrium geometry. The x-axis represents imaginary time, over which the AD is performed, starting with equilibrated walkers. After around 8 a.u., the dipole moment converges to a value outside the stochastic error bar of the finite difference AFQMC estimator calculated using common random numbers. This discrepancy arises from the bias due to SR in sampling AD derivatives, leading to a bias of 0.003(1) a.u in this case. Since this is a relatively small system, we can perform AD-AFQMC calculations without SR and this value matches with the finite difference calculation within stochastic error. However, without SR, more sampling effort is required to achieve the same level of stochastic error. One can perform SR less frequently to control this bias, but we do not explore this approach further since the bias is relatively small. We also show the converged value of the AD-AFQMC dipole moment (with SR) without including trial orbital relaxation, illustrating that including orbital relaxation brings the dipole moment closer to the exact value.

In the bottom panel of figure \ref{fig:conv}, we conduct a similar analysis for the one-electron energy of a twenty-atom linear hydrogen chain with an interatomic separation of 2.4 a.u. in the minimal STO-6G basis. We use an unrestricted HF (UHF) trial state for the AFQMC calculations. In this case, the SR bias of AD-AFQMC is 0.001(1) a.u. The orbital relaxation contribution is substantial and increases the accuracy of the estimator significantly. 

\subsection{Hydrogen chains}\label{sec:hchains}
\begin{figure*}[t]
   \includegraphics[width=1.8\columnwidth]{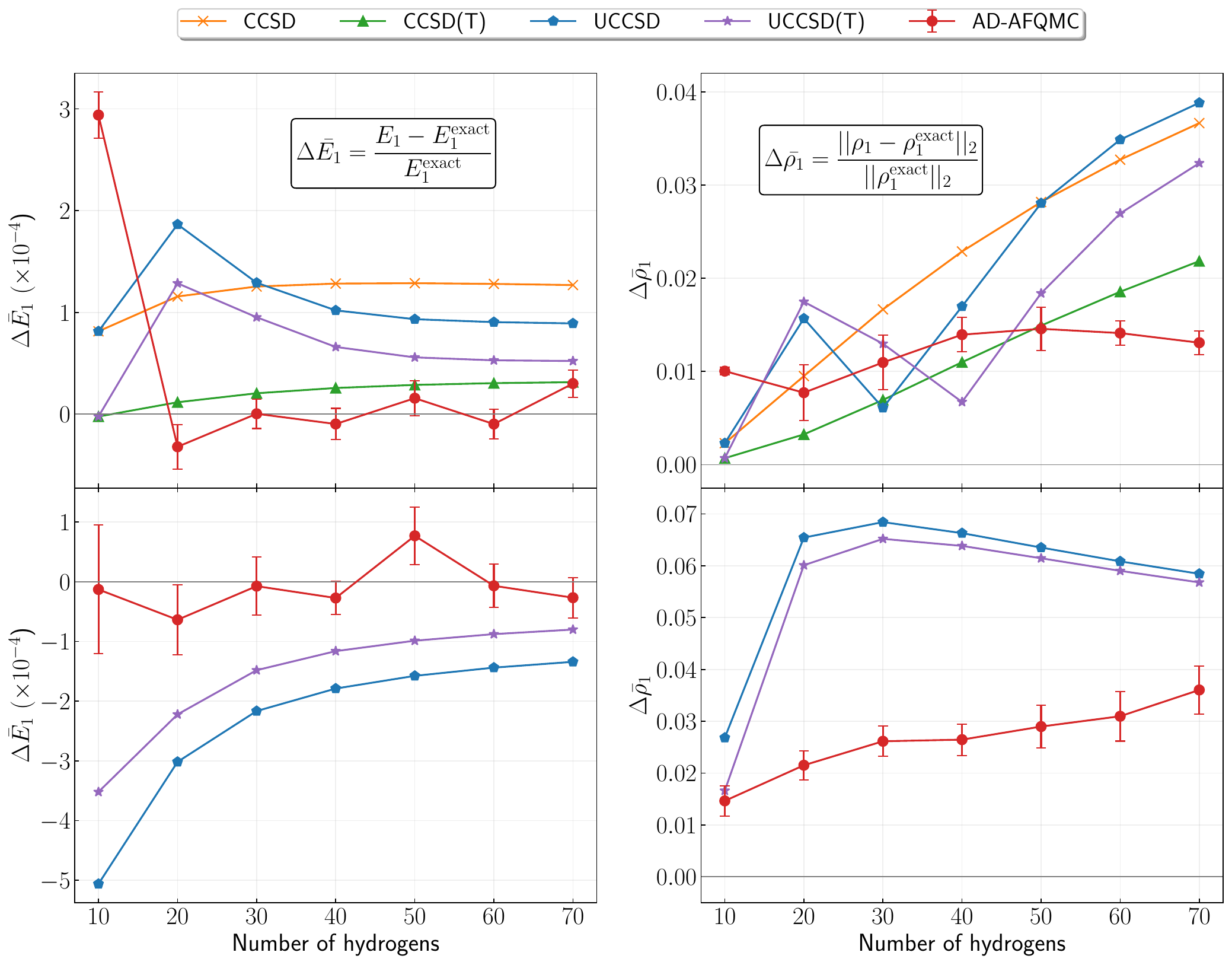}
   \caption{Relative error in the one-electron energy (left) and 1-RDM (right) for a hydrogen chain (STO-6G basis) with an inter-hydrogen separation of \(d=1.6\) Bohr (top) and \(d=2.4\) Bohr (bottom).}\label{fig:hchains}
\end{figure*}
In figure \ref{fig:hchains}, we present a comparison of the accuracy of AD-AFQMC one-electron energies and 1-RDMs with coupled cluster methods for hydrogen chains. Near-exact reference benchmark results for this one-dimensional system were obtained using DMRG. Unrestricted coupled cluster calculations were performed on top of UHF references. We did not include orbital relaxation in CC calculations, as we found it not to significantly impact the results. All AFQMC calculations were done with UHF trial states and with SR. 1-RDM errors were calculated using the Frobenius norm. We consider two bond lengths (\(d\)) of 1.6 a.u. and 2.4 a.u. 

For \(d=1.6\) a.u., out of the CC methods CCSD(T) performs very well, as one would expect for a system close to equilibrium. We note that for the same bond length, the chains become more correlated with increasing length, explaining the increasing error in both the one-electron energy and 1-RDM for the CC methods. AD-AFQMC one-electron energy systematic errors are smaller than CCSD(T) for all chain lengths, except for \ce{H10}. This discrepancy is due to the proximity of \(d=1.6\) a.u. to the Coulson-Fischer point of this chain, beyond which the UHF solution becomes more stable. The AD-AFQMC 1-RDM error is also smaller than CCSD for all chains except \ce{H10} and smaller than CCSD(T) beyond \ce{H50}. 

For \(d=2.4\) a.u., the CC methods are less accurate due to the increased correlation in the system. In this case, restricted methods have convergence issues, so we only present the results for unrestricted observables. For both one-electron energies and 1-RDMs, AD-AFQMC errors are significantly smaller compared to UCCSD and UCCSD(T), demonstrating the efficacy of AD-AFQMC in providing accurate estimates of properties across a range of correlation strengths.

\begin{figure}
   \includegraphics[width=\columnwidth]{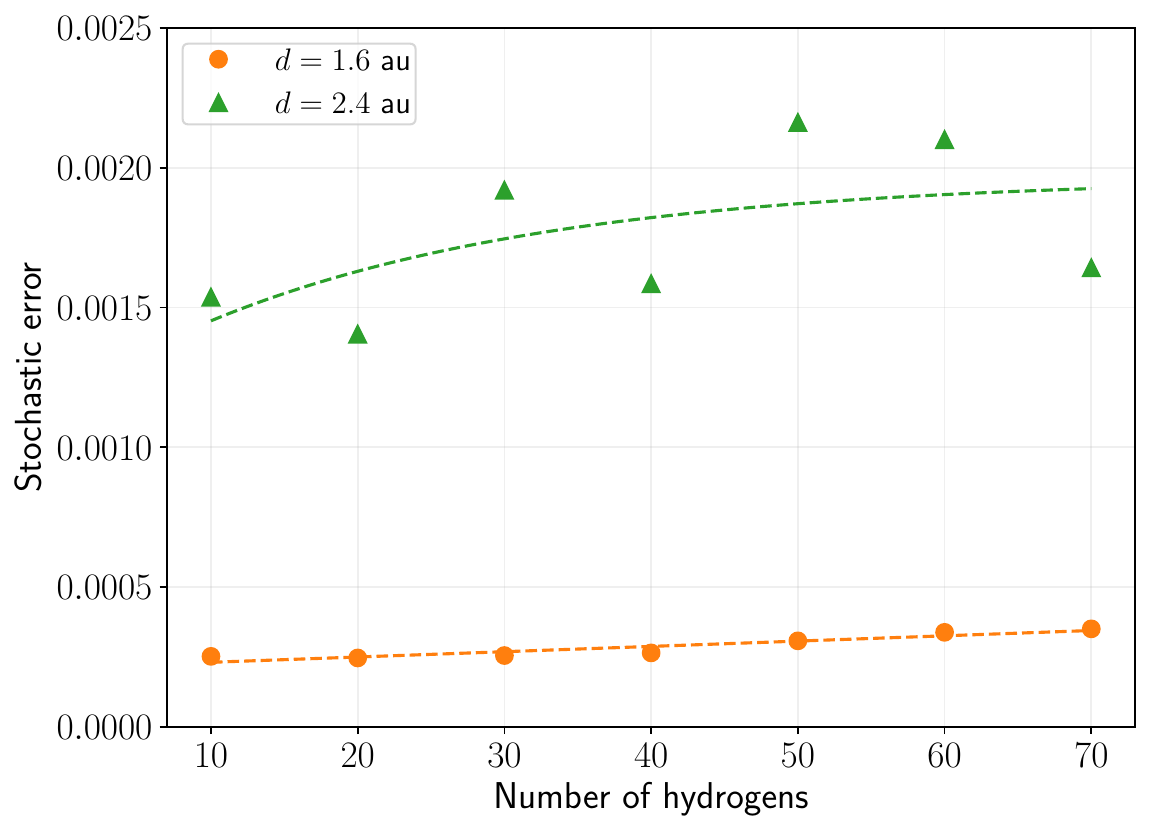}
   \caption{Stochastic error in the occupation number in an atomic orbital in the middle of a hydrogen chain (STO-6G basis).}\label{fig:occ}
\end{figure}

Analyzing the system-size scaling of stochastic errors in AD-AFQMC estimates is crucial to assess their applicability in large systems. For a local observable, it is desirable that the stochastic error not grow with system size, since the observable itself does not scale. In figure \ref{fig:occ}, we show the stochastic error in the occupation of a hydrogen 1\(s\) STO-6G atomic orbital in the middle of a hydrogen chain for \(d=1.6\) a.u. and \(d=2.4\) a.u. for the same amount of sampling effort. Remarkably, it is observed that the stochastic error does not increase with system size asymptotically. While the stochastic error in energy increases as \(\sqrt{N_{\text{Hydrogens}}}\), the noise in the energy derivative does not increase. This behavior has previously been reported for correlated sampling estimators in \citenum{assaraf2011chaotic}.

\begin{figure}
   \includegraphics[width=\columnwidth]{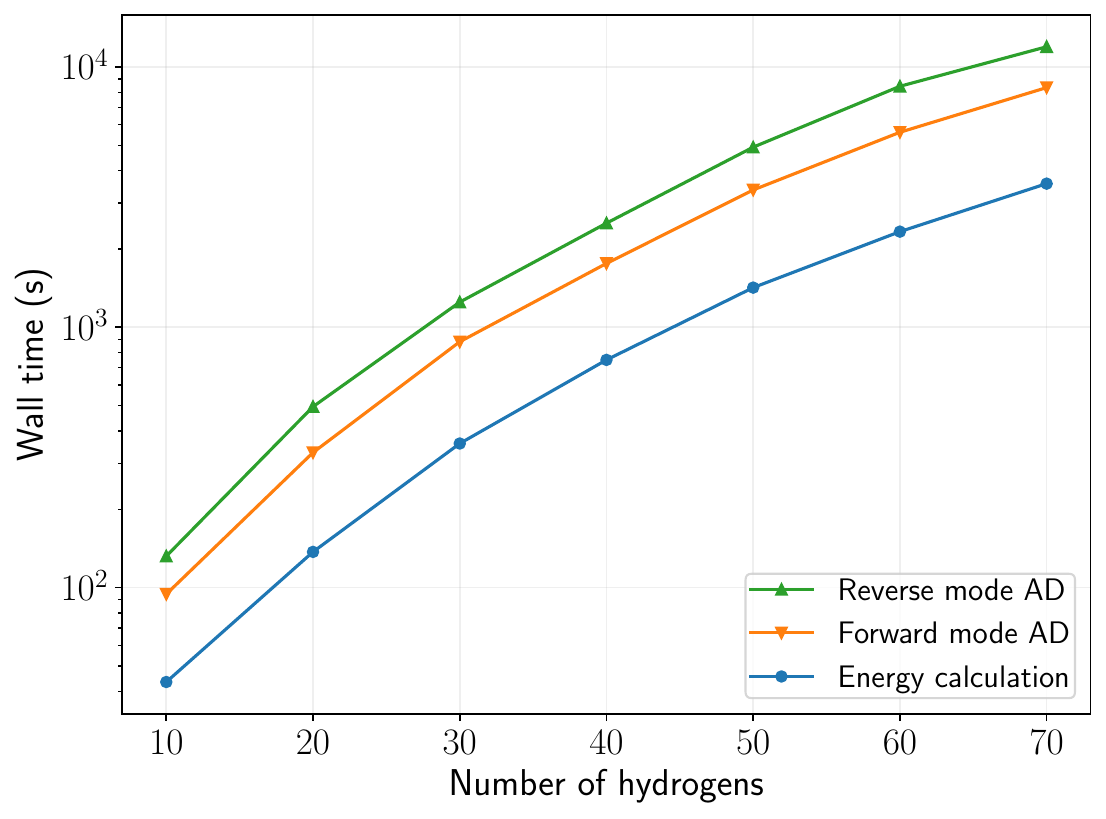}
   \caption{Computational time scaling of AD-AFQMC calculations for hydrogen chains of different lengths with the same sampling effort. Reverse mode AD times correspond to the 1-RDM calculations shown in figure \ref{fig:hchains}, while forward mode AD times correspond to the occupation number calculations in figure \ref{fig:occ}.}\label{fig:times}
\end{figure}

In figure \ref{fig:times}, we show the cost (wall time, including compilation time) of forward and reverse mode AD-AFQMC 1-RDM calculations relative to that of energy. As expected, the reverse mode AD costs are 3 to 4 times as expensive as the energy calculation for all chain lengths, while the forward mode AD costs are 2 to 3 times as expensive. We also show the size scaling of energy calculation cost, which is seen to be between quadratic and cubic. We note that computational cost scaling of the backpropagation algorithm for the calculation of properties and forces is also similarly favorable.\cite{motta2017computation,motta2018communication}

\subsection{Dipole moments}\label{sec:dipoles}

\begin{table*}
   \caption{Dipole moments (in a.u.) of molecules at equilibrium geometries using the aug-cc-pVTZ basis set. BP-AFQMC refers to estimates obtained using backpropagation in AFQMC with partial path restoration.}\label{tab:dipoles}
   \centering
   \begin{tabular}{*{13}c}
   \hline
   Molecule &~~& MP2 &~~& CCSD &~~& BP-AFQMC &~~& AD-AFQMC &~~& CCSD(T) &~~& Experiment \\
   \hline
   \ce{H2O} && 0.7298 && 0.7335 && 0.707(4) && 0.720(2) && 0.7247 && 0.730(2)\cite{lide2004crc}\\
   \ce{NH3} && 0.5996 && 0.6015 && 0.578(6) && 0.592(2) && 0.5938 && 0.581(1)\cite{shimizu1970stark} \\
   \ce{HCl} && 0.4389 && 0.4318 && 0.400(5) && 0.429(1) && 0.4273 && 0.430\cite{lovas2005diatomic} \\
   \ce{HBr} && 0.3379 && 0.3289 && 0.33(1) && 0.329(2) && 0.3245 && 0.325\cite{lovas2005diatomic} \\
   \ce{CO} && 0.1050 && 0.0199 && 0.12(1) && 0.019(4) && 0.0429 && 0.048(1)\cite{meerts1977electric}\\
   \ce{CH2O} && 0.9375 && 0.9666 && 0.82(1) && 0.965(5) && 0.9389 && 0.918(1)\cite{theule2003fluorescence}\\
   \ce{C4H5N} && 0.7354 && 0.7241 && - && 0.730(9) && 0.7255 && 0.72(2)\cite{nelson1967selected}\\
   \hline
   \end{tabular}
\end{table*}

In table \ref{tab:dipoles}, we show a comparison of dipole moments calculated using AD-AFQMC with those calculated using backpropagation (BP-AFQMC) and with conventional quantum chemistry methods. We use the aug-cc-pVTZ basis which is large enough to allow meaningful comparison with experimental values. The equilibrium geometries used here are provided in reference \citenum{afqmc_files}. For pyrrole, we used the geometry reported in reference \citenum{loos2020mountaineering}. We include orbital relaxation contributions and use the frozen-core approximation in all methods. Orbital relaxation contributions are not relevant for BP-AFQMC. 

Since these molecules are largely single reference, CCSD(T) values are expected to be very accurate. We see that AD-AFQMC dipole moments are in better agreement with CCSD(T) values than CCSD in most cases. They are also close to experimental dipole moments with some residual basis set error. This is an encouraging sign considering that AFQMC has a less steep cost scaling compared to the CC methods. Results obtained with BP-AFQMC using partial path restoration are significantly less accurate than AD-AFQMC for \ce{CO} and \ce{CH2O}. While the sampling effort in BP and AD-AFQMC is not directly comparable, we used a similar allocation of computational resources for both sets of calculations. Note that one can use better trial states to obtain improved BP estimates as demonstrated in reference \citenum{motta2017computation}. Details about the convergence of BP-AFQMC are provided in the supporting information.

\begin{figure}[htp]
   \includegraphics[width=\columnwidth]{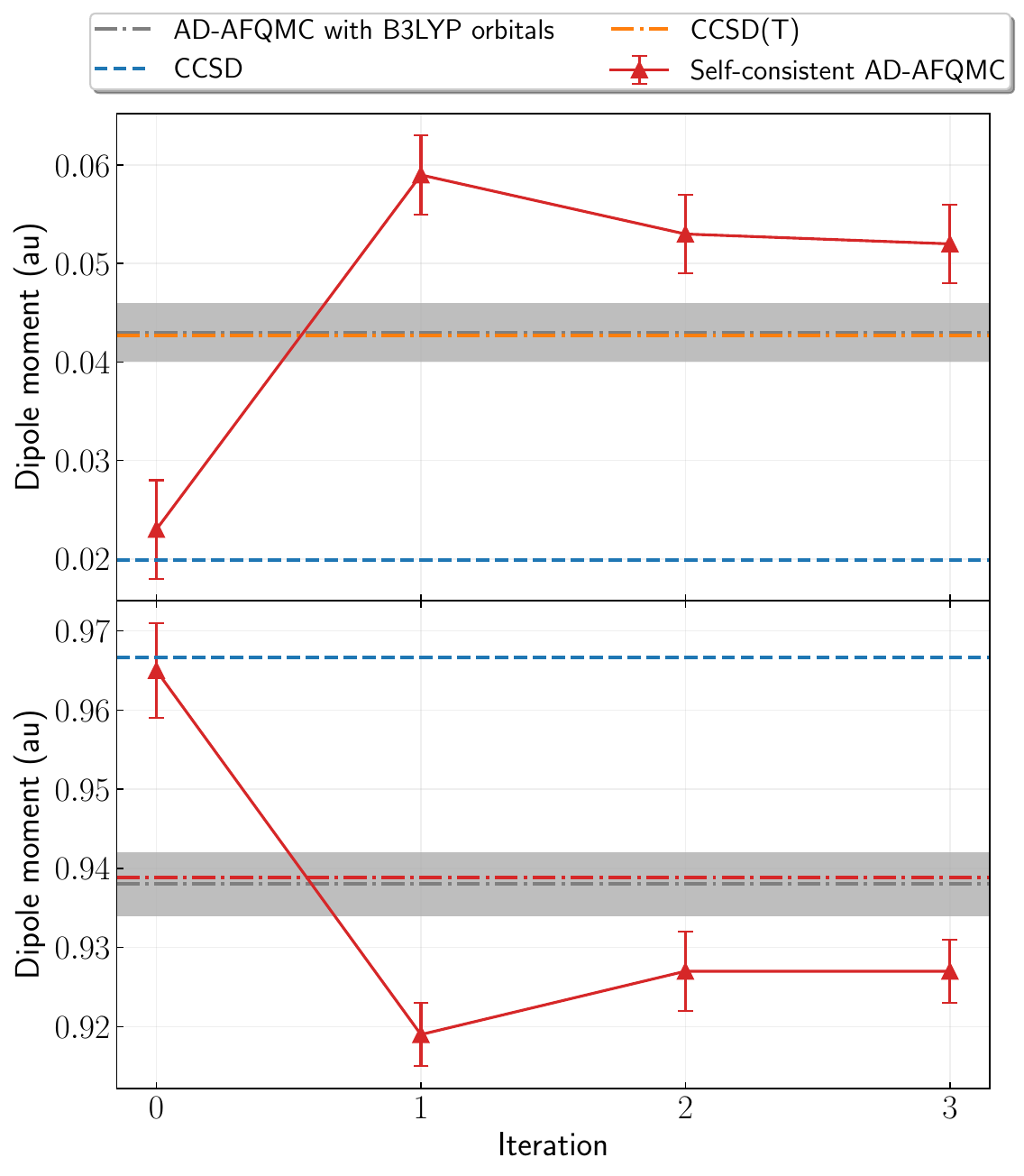}
   \caption{Convergence of the dipole moment of \ce{CO} (top panel) and \ce{CH2O} (bottom panel) with the number of self-consistent AFQMC iterations using the aug-ccpVTZ basis.
   Note that self-consistency performs worse than using B3LYP orbitals.}\label{fig:sc_dipole}
\end{figure}

We see greater errors in both AD-AFQMC and CCSD dipole moments of \ce{CO} and \ce{CH2O} molecules. To mitigate this error, we use the self-consistent AFQMC approach.\cite{qin2016coupling,qin2023self} We calculate the 1-RDM using reverse-mode AD-AFQMC and use the corresponding natural orbitals in the trial state of the subsequent AFQMC calculation. This procedure is repeated until the desired convergence threshold. Note that we include orbital relaxation contributions in the zeroth iteration to obtain an accurate 1-RDM initially, but not in the subsequent iterations. The results of this procedure for \ce{CO} and \ce{CH2O} are shown in figure \ref{fig:sc_dipole}. In both cases, the self-consistent iterations can be seen to improve the dipole accuracy with convergence achieved in less than five steps. Availability of 1-RDM from AD-AFQMC can thus enable systematic improvement in AFQMC properties. We also show the results of a single AD-AFQMC calculation using DFT-B3LYP as the trial state without orbital relaxation. These dipole moments show better agreement with the CCSD(T) value suggesting that these orbitals are optimal for AFQMC calculations. Thus it may be possible to devise orbital optimization schemes other than the one used in self-consistent AFQMC to obtain better trial states for property calculations.   

\section{Conclusion}\label{sec:conclusion}
In this work, we investigated the utility of automatic differentiation (AD) for calculating response properties in auxiliary field quantum Monte Carlo (AFQMC) with the aim of achieving accuracy and computational cost comparable with energy calculations. We studied various contributions to energy derivatives and their convergence behavior. Our analysis demonstrated the accuracy of AD-AFQMC for computing one-particle reduced density matrices (1-RDM) in hydrogen chains, achieved at a computational cost no more than four times that of the energy calculation. We found that the stochastic error of the AD-AFQMC estimator for a local orbital density does not increase with system size. Finally, dipole moments of several molecules calculated using AD-AFQMC were found to be in good agreement with CCSD(T) and experiments. We were able to further improve the accuracy of these dipole moments by employing self-consistent AFQMC calculations, and alternatively by using trial states from DFT.

A way to improve AD-AFQMC involves addressing the bias due to stochastic reconfiguration. While for the systems studied in this paper, the SR bias was found to be small, it is possible that in systems where there are large fluctuations in the weights, this bias can be significant. We plan to explore bias mitigation techniques in future work. Another avenue for improvement lies in using more accurate trial states. In this work, we used a single determinant trial state, but it is possible to use multi-determinant trial states. Furthermore, to enable calculations on much larger systems, advancements in AD implementations are necessary. Developing more sophisticated checkpointing schemes may help manage the memory usage of reverse-mode AD, allowing for more extensive calculations without excessive memory constraints. Overall, these potential enhancements will contribute to the broader applicability of AD-AFQMC.

Despite these challenges, we believe that
 AD-AFQMC presents a viable and promising approach for calculating properties in AFQMC. A natural extension of this work is the calculation of nuclear gradients using AD-AFQMC, which can be accomplished by combining our code with differentiable molecular integral evaluators. The low cost-scaling and high-accuracy of AD-AFQMC open up possibilities for extremely accurate molecular dynamics simulations.

 \section*{Data availability}
Code used for AD-AFQMC calculations is available in a public GitHub repository at reference \citenum{dqmc_code}, and the input and output files for these calculations are also available at reference \citenum{afqmc_files}.

 \section*{Acknowledgements}
We would like to thank Matthew Foulkes, Shiwei Zhang, and Miguel Morales for helpful discussions. AM and JSK were supported by NSF Career award CHE-2145209. SS was supported by a grant from the Camille and Henry Dreyfus Foundation. DRR acknowledges funding support from NSF CHE--2245592. This work utilized resources from the University of Colorado Boulder Research Computing Group, which is supported by the National Science Foundation (Award Nos. ACI-1532235 and ACI-1532236), the University of Colorado Boulder, and Colorado State University. Some calculations were also performed at the Advanced Research Computing at Hopkins (ARCH) core facility through allocation CHE230028 from the Advanced Cyberinfrastructure Coordination Ecosystem: Services and Support (ACCESS) program, which is supported by National Science Foundation grants \#2138259, \#2138286, \#2138307, \#2137603, and \#2138296. 

\providecommand{\latin}[1]{#1}
\makeatletter
\providecommand{\doi}
  {\begingroup\let\do\@makeother\dospecials
  \catcode`\{=1 \catcode`\}=2 \doi@aux}
\providecommand{\doi@aux}[1]{\endgroup\texttt{#1}}
\makeatother
\providecommand*\mcitethebibliography{\thebibliography}
\csname @ifundefined\endcsname{endmcitethebibliography}
  {\let\endmcitethebibliography\endthebibliography}{}

\end{document}